\documentclass[11pt,a4paper]{article}

\usepackage[english]{babel}
\usepackage[utf8]{inputenc}

\usepackage{epsf}
\usepackage{cite}
\usepackage{graphicx}
\usepackage{subcaption} 

\usepackage{amsmath}
\usepackage{amssymb}
\usepackage{mathrsfs}
\usepackage[normalem]{ulem}
\usepackage{mathbbol}
\usepackage{multirow}
\usepackage{slashed}
\usepackage{amsfonts}
\usepackage{bbding}
\usepackage{array}
\usepackage[
colorlinks=true
,urlcolor=black
,anchorcolor=black
,citecolor=black
,filecolor=black
,linkcolor=black
,menucolor=black
,linktocpage=true
,pdfproducer=medialab
,pdfa=true
]{hyperref}

\usepackage{color} 
\usepackage[left=2cm,right=2cm,top=2cm,bottom=2cm]{geometry}
\usepackage{tikz}
\usetikzlibrary{shapes,arrows,positioning,automata,backgrounds,calc,er,patterns}
\usepackage[compat=1.1.0]{tikz-feynman}
\usepackage{contour}
\usepackage{placeins}
\allowdisplaybreaks
\begin{document}

\begin{center}
{\Large\bf
Exploring asymmetries in three-body cLFV lepton
decays: \vspace*{3mm}\\
probing CP violation in HNL extensions of the SM
}

\vspace*{1.2cm}
{\bf Adrian Darricau\footnote{Poster presented by Adrian Darricau at cLFV 2026, Rome, 17 - 19 june 2026.} $^\text{a}$, 
Jonathan~Kriewald $^\text{b}$,  Ana~M.~Teixeira $^\text{a}$ }

\vspace*{.5cm}
$^\text{a}$ Laboratoire de Physique de Clermont Auvergne (UMR 6533), CNRS/IN2P3,\\
Univ. Clermont Auvergne, 4 Av. Blaise Pascal, 63178 Aubi\`ere Cedex,
France
 
\vspace*{.2cm}
$^\text{b}$ {Jožef Stefan Institut, Jamova Cesta 39, P. O. Box 3000, 1001 Ljubljana, Slovenia}
\end{center}

\vspace*{5mm}
\begin{abstract}
\noindent
In extensions of the Standard Model with Majorana sterile fermions, additional Dirac and Majorana CP-violating phases can have a significant impact on charged lepton flavour violating (cLFV) transitions and decays. 
We investigate angular observables associated with polarised $\mu$ and $\tau$ cLFV three-body decays, $\ell_\alpha^+ \to \ell_\beta^+ \ell_\gamma^+ \ell_\delta^-$, including parity- and time-reversal asymmetries. We analyse the correlations between different cLFV observables, showing that sizeable asymmetries can arise over experimentally accessible regions of parameter space. 
The resulting patterns of angular observables provide powerful probes of this class of Standard Model extensions and, in the event of a cLFV signal, could offer a means to test or potentially falsify the underlying framework.

\end{abstract}

\section{Introduction}

Neutrino oscillations provide compelling evidence that the Standard Model (SM) is incomplete, establishing that neutrinos are massive and leptons undergo flavour transitions. This naturally raises the possibility of charged lepton flavour violation (cLFV); moreover the now present leptonic mixings open the door to new sources of leptonic CP violation. Among the many scenarios proposed to account for neutrino masses, extensions of the SM via (Majorana) sterile fermions stand out as some of the simplest and yet most attractive possibilities.

Depending on the mechanism of neutrino mass generation and on the scale at which it is realised, the expected phenomenological implications vary significantly. Low-scale extensions of the SM via heavy neutral leptons (HNL), with masses around the TeV scale and non-negligible mixings with the active neutrinos, are particularly appealing, as they can induce sizeable contributions to charged lepton flavour violating (cLFV), lepton number violating (LNV) and CP violating (CPV) observables, while also providing new sources of CP violation relevant for leptogenesis. In the CP-conserving limit, several interesting correlations between observables have been identified, providing a means to falsify these constructions in the advent of (multiple) experimental measurements.

In general, SM extensions with HNL contain additional Dirac and Majorana CP-violating phases. Besides their potential role in leptogenesis, these phases can significantly affect cLFV observables by modifying total rates and spoiling the characteristic correlation patterns otherwise found in the CP-conserving limit~\cite{Abada:2021zcm}.

In the presence of generic CP-violating phases, the standard set of cLFV observables ($\ell_\alpha \to \ell_\beta \gamma$, $\ell_\alpha \to \ell_\beta \ell_\gamma \ell_\delta$, neutrinoless $\mu-e$ conversion, ...) is no longer sufficient to disentangle the underlying operator contributions. Motivated by the high degree of polarisation of muon beams at PSI (Mu3e experiment), the work carried out in~\cite{Darricau:2025rmu} investigated asymmetries in the angular distributions of polarised $\ell_\alpha^+\to\ell_\beta^+\ell_\gamma^+\ell_\delta^-$ decays, showing that they provide a powerful probe of the underlying Wilson coefficients in the general CP-violating regime.

\section{The simplified ``3+2'' HNL framework}
We rely on a minimal ``ad-hoc'' extension of the SM via two generations of (Majorana) heavy sterile states, the so-called ``3+2'', which can be regarded as a simplified description of low-scale seesaw scenarios. The neutral spectrum is composed of 5 Majorana states: 3 light (mostly active) neutrinos, and two additional heavy ones, with masses $m_{4,5}$. It is important to clarify that no assumption is made on the underlying mechanism at the origin of neutrino masses; one hypothesises that leptonic mixings are encoded in an enlarged $5\times5$ unitary mixing matrix, $\mathcal U$. 
The upper left $3\times3$ block of the latter thus corresponds to the left-handed leptonic mixing matrix, the would-be PMNS, $\tilde{U}_\text{PMNS}$.
Under such an ad-hoc approach, $\mathcal{U}$ can be parametrised through five subsequent rotations $R_{ij}$ (with $i\neq j$), and a diagonal matrix which encodes the four physical Majorana phases, $\varphi_i$
\begin{eqnarray}
    \mathcal{U} \,= \,R_{45}\,R_{35}\,R_{25}\,R_{15}\,R_{34}\,R_{24}\,R_{14}\,R_{23}\,R_{13}\,R_{12}\times\mathrm{diag}(1, e^{i\varphi_2}, e^{i\varphi_3}, e^{i\varphi_4}, e^{i\varphi_5})\,.
    \label{eqn:allrot}
\end{eqnarray}
In the presence of non-vanishing active-sterile mixings (parametrised via $\theta_{\alpha 4}$ and $\theta_{\alpha 5}$), the would-be PMNS mixing matrix will be non-unitary; it proves convenient to parametrise the deviations from unitarity via a matrix $\eta$
\begin{equation}
\label{eq:defPMNSeta}
U_\text{PMNS} \, \to \, \tilde U_\text{PMNS} \, = \,(\mathbb{1} - \eta)\, 
U_\text{PMNS}\,.
\end{equation}
The deviations from unitarity of the would-be PMNS mixing matrix induce modified charged- and neutral-current interactions, affecting electroweak precision observables (EWPO), lepton flavour universality (LFU) tests and cLFV processes. We impose the generic bounds on $\eta$, including EWPO, LFU tests and measurements of $G_F$, $\sin^2 \theta_w$ and the mass of the $W$-boson among others.

Moreover, the presence of the new Majorana heavy states can also lead to modifications of the predictions regarding neutrinoless double beta decays, in particular of the effective mass $m_{ee}$. We further impose the KamLAND-ZEN bound on the effective Majorana mass $m_{ee}^\mathrm{eff}$. 

The light-neutrino parameters are fixed within the $3\sigma$ NuFIT~6.0 ranges assuming normal ordering and a fixed lightest neutrino mass scale at $10^{-4}$~eV.

Lastly, the most important bounds to be considered in this study in general arise from current searches for rare cLFV transitions and decays, which lead to stringent constraints on HNL masses and mixing parameters. In addition to the cLFV three-body decays (which are at the core of the present study), the most relevant processes include radiative decays, $\mu - e$ conversion in atoms, and cLFV $Z$ boson decays.

\section{Kinematics and angular asymmetries in three-body decays}
\begin{figure}[h!]
    \centering
    \includegraphics[width=0.45 \textwidth] {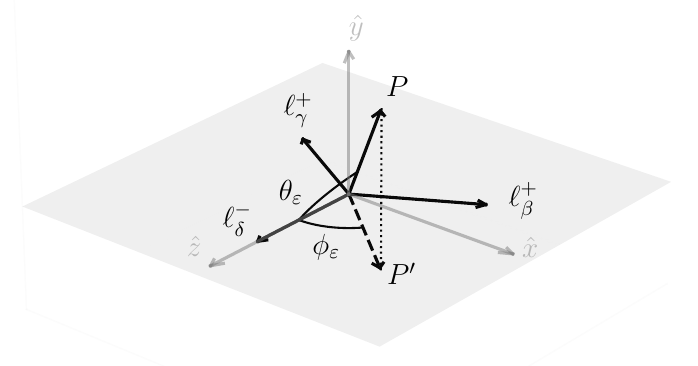}
    \caption{Rest frame of the decaying charged lepton $\ell_\alpha^+$, with $\vec{P}$ denoting its polarisation. }
    \label{fig:ref-frame}
\end{figure}

To define the angular asymmetries, we first introduce the kinematics of the generic pure cLFV leptonic three-body decay
\begin{equation}\label{eq:genericdecay}
    \ell_\alpha^+ (p) \to 
    \ell_\beta^+ (k_1)\, \ell_\gamma^+ (k_2)\, \ell_\delta^-(k_3)\,.
\end{equation}
As schematically shown in Fig.~\ref{fig:ref-frame} and following~\cite{Bolton:2022lrg}, we work in the rest frame of the decaying lepton. In this frame, the angle $\theta_\varepsilon$ is defined as the angle between the decaying lepton polarisation vector and the outgoing negatively charged lepton, while $\phi_\varepsilon$ denotes the azimuthal angle in the decay plane.
Denoting the polarisation four-vector of the decaying lepton as $(0, \vec{P})$, one has the following expression for the double-differential decay width, decomposed in spin-independent and spin-dependent part
\begin{align}\label{eq:DoubleDiffDec}
    \frac{d \Gamma_{\ell_\alpha^+ \to \ell_\gamma^+ \ell_\beta^+ \ell_\beta^-}}{d \Omega_\epsilon} &= \frac{\Gamma_{\ell_\alpha^+ \to \ell_\gamma^+ \ell_\beta^+ \ell_\beta^-}}{4 \pi} \left[1 + P\left(\mathcal{A}^P_{\ell_\alpha^+ \to \ell_\gamma^+ \ell_\beta^+ \ell_\beta^-}  \cos \theta_\varepsilon \right.\right. \nonumber \\
    &+ \left. \left.\mathcal{A}^{P^\prime}_{\ell_\alpha^+ \to \ell_\gamma^+ \ell_\beta^+ \ell_\beta^-}  \sin \theta_\varepsilon \cos \phi_\varepsilon + \mathcal{A}^T_{\ell_\alpha^+ \to \ell_\gamma^+ \ell_\beta^+ \ell_\beta^-}  \sin \theta_\varepsilon \sin \phi_\varepsilon\right) \right]\,,
\end{align}
with $P \equiv \vert \vec{P} \vert$. While the total decay rate can be obtained by integrating over the entire phase space, the angular coefficients $\mathcal A_{P,P',T}$ can be projected out via asymmetric integrations over the angles. We note here that in the case of same-flavour final states the sign of the azimuthal angle $\phi_\varepsilon$ is not well defined. Without loss of generality (and without loss of phase space), we further restrict the integration region with an energy ordering of the same-charge same-flavour final states as $E_1 \gtrless E_2$ (either ordering being acceptable). In the case of opposite-flavour final states, the full integration region must be taken.
The energy ordering can be used to further introduce ``forward-backward'' asymmetries by  defining an asymmetric integration region in the Dalitz plane as
\begin{align}\label{eq:Omega:def}
    \int_{\Omega(s_1,s_2)} \to \left(\int_{E_1 > E_2} - \int_{E_1 < E_2}\right)\,.
\end{align}
This allows introducing new observables: $\mathcal A_{FB},\,\mathcal{A}_{FB,P},\,\mathcal A_{FB,P'}$ and $\mathcal A_{FB,T}$.

\section{Prospective study of the asymmetries in cLFV three-body decays}\label{sec:full-results}

In what follows, we focus on $\mu$ and $\tau$ three-body cLFV decays and illustrate the interplay between angular asymmetries and cLFV rates.
Experiments dedicated to searches for cLFV muon transitions are expected to significantly improve their sensitivity in the near future, either strengthening existing bounds, or hopefully observing a SM-forbidden processes. It is thus interesting to assess the potential for simultaneous observation of several cLFV quantities: decay (conversion) rates and asymmetries. As shown in~\cite{Abada:2021zcm}, CP-violating phases can significantly disrupt the correlation patterns expected among cLFV observables. This makes it essential to consider a broader set of observables in order to assess whether cLFV originates from HNL mixing.

\begin{figure}[h!]
\centering
\includegraphics[width=0.48 \textwidth] {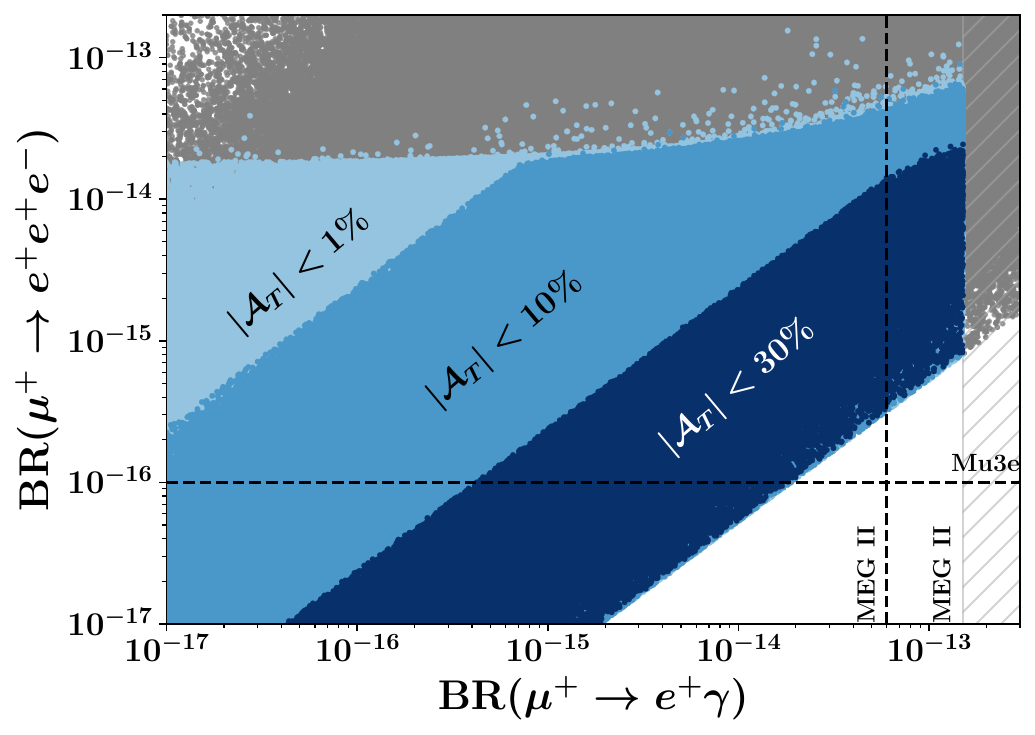}\hspace*{8mm}
\includegraphics[width=0.48 \textwidth] {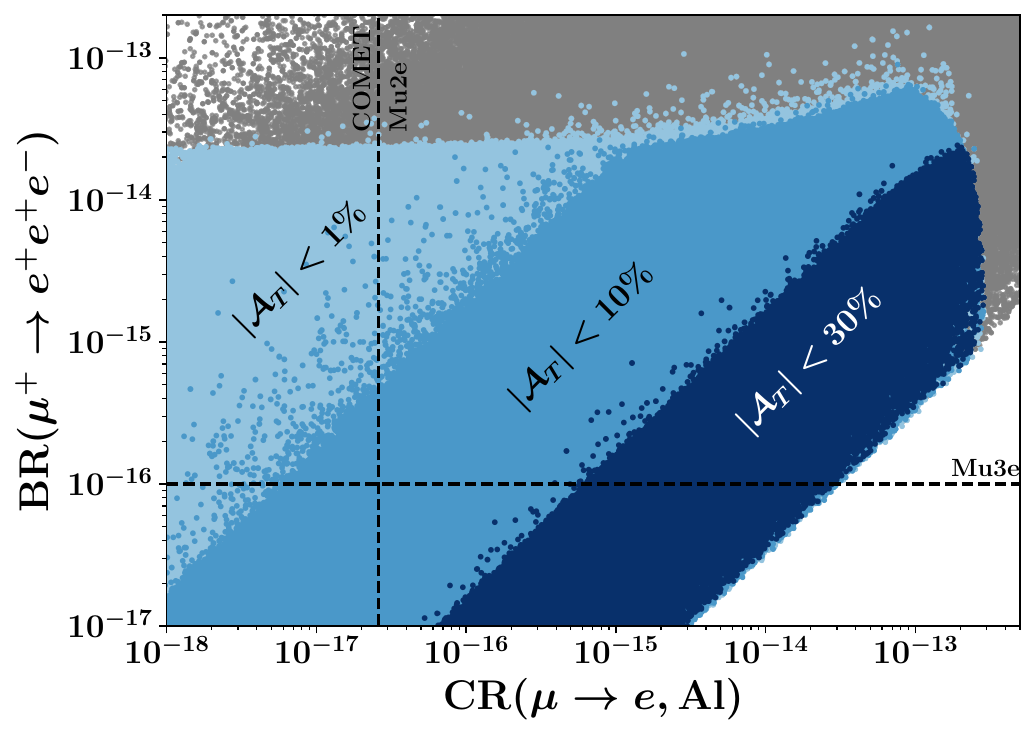}
\caption{Projected values of the $T$-asymmetry in $\mu \to 3e$ decays. Left: BR($\mu \to 3e$) vs. BR($\mu \to e\gamma$). Right: BR($\mu \to 3e$) vs. CR($\mu$–$e$, Al). Blue bands indicate the maximal attainable values of $|\mathcal{A}^T|$. Grey (and grey-dashed) regions are excluded by current bounds. Horizontal and vertical lines indicate current limits and future sensitivities.}
\label{fig:Mue_TAsym}
\end{figure}

In Fig.~\ref{fig:Mue_TAsym}, we show the expected ranges of the $T$-asymmetry in $\mu \to 3e$ decays in the planes spanned by BR($\mu \to 3e$) and either BR($\mu \to e\gamma$) or the $\mu$–$e$ conversion rate in nuclei.
As shown in both panels of Fig.~\ref{fig:Mue_TAsym}, values of the $T$-asymmetry as large as $\sim 25\%$ can be obtained in regions accessible to future muon–electron cLFV searches. The diagonal structure of the bands reflects the interplay between dipole contributions and four-lepton vector operators, in particular $Z$-penguin terms.

The combined measurement of the three cLFV rates and of the $T$-asymmetry provides a powerful probe of this class of models. For instance, a future observation of BR($\mu \to 3e$)$\sim \mathcal{O}(10^{-15})$ together with $\mathcal{A}^T \sim \mathcal{O}(20\%)$ would imply a $\mu$–$e$ conversion rate in aluminium above $10^{-14}$.

\begin{figure}[h!]
\centering
\includegraphics[width=0.48 \textwidth] {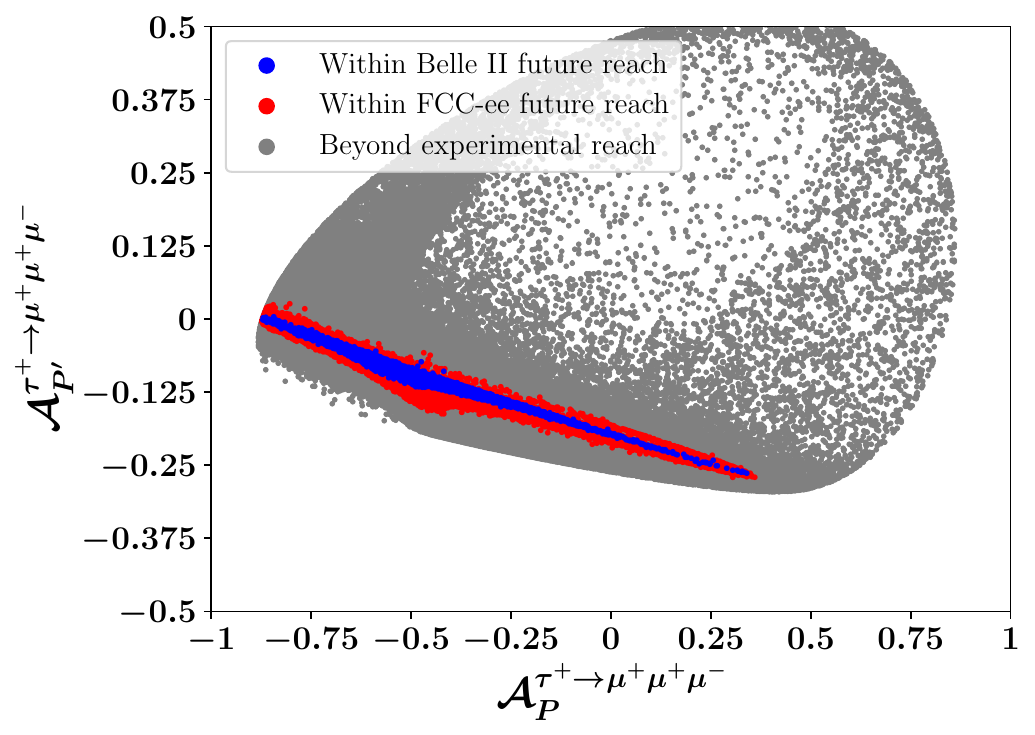}
\hspace*{3mm}
\includegraphics[width=0.48 \textwidth] {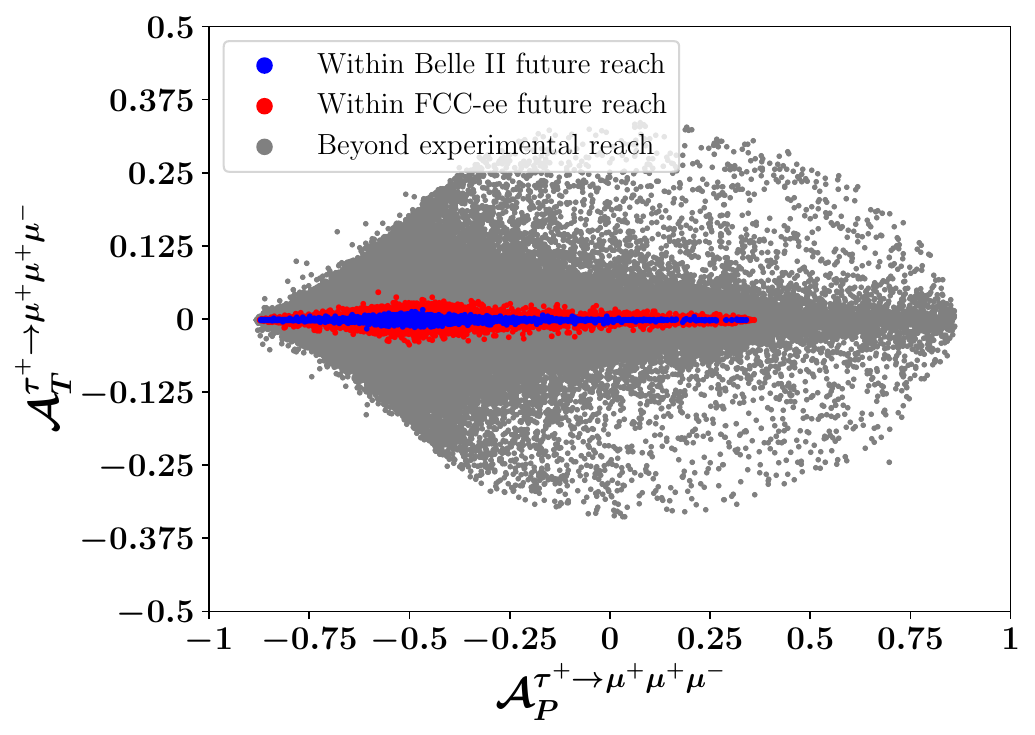}
\caption{Joint prospects for $\tau \to 3 \mu$ asymmetries. 
All points displayed are in agreement with current experimental bounds; the blue (grey) points are within (beyond) future sensitivity for the decays under consideration. 
Blue points correspond to Belle II's future reach, and red to FCC-ee's expected sensitivity.}
\label{fig:l3l_Asym_vs_Asym}
\end{figure}

The situation is markedly different for $\tau \to 3\mu$ decays. As shown in Fig.~\ref{fig:l3l_Asym_vs_Asym}, the expected sensitivity of Belle II (and FCC-ee) probes only a restricted region of the asymmetry parameter space, which is associated with regimes dominated by $Z$-penguin contributions. Nonetheless, the limited experimental reach in the $\tau$ sector still allows for meaningful tests of this minimal HNL framework, since the $P$, $P^\prime$ and $T$ asymmetries in $\tau \to 3\mu$ channels remain strongly correlated and confined within relatively narrow ranges.

\begin{figure}[h!]
\centering
\includegraphics[width=0.48 \textwidth] {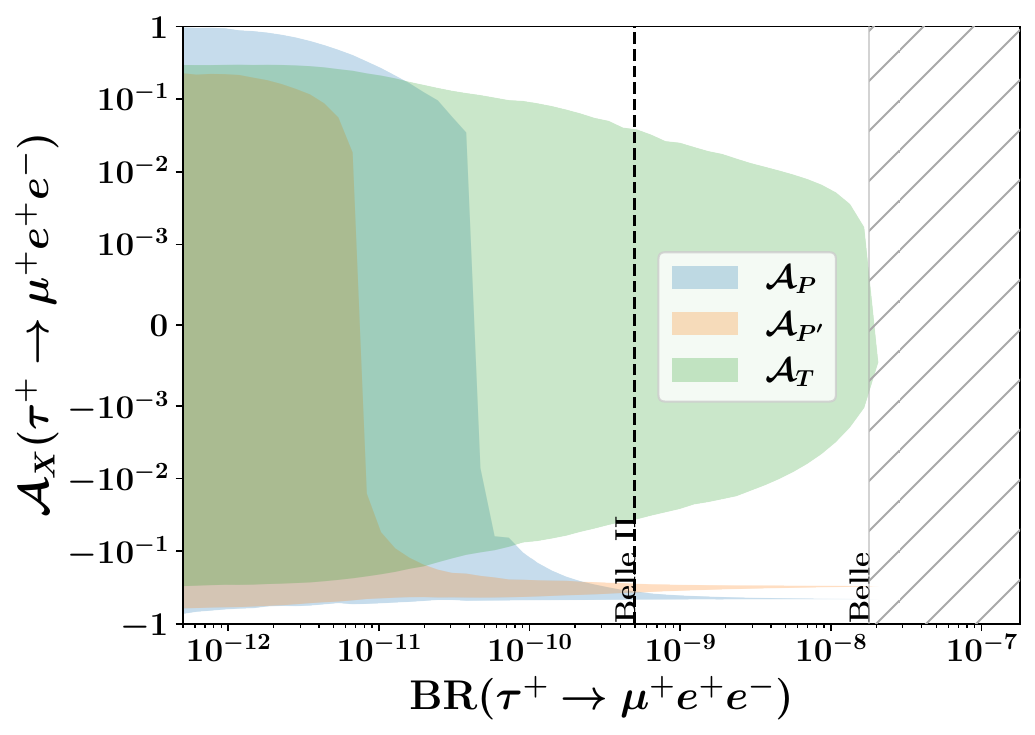}
\hspace*{3mm}
\includegraphics[width=0.48 \textwidth] {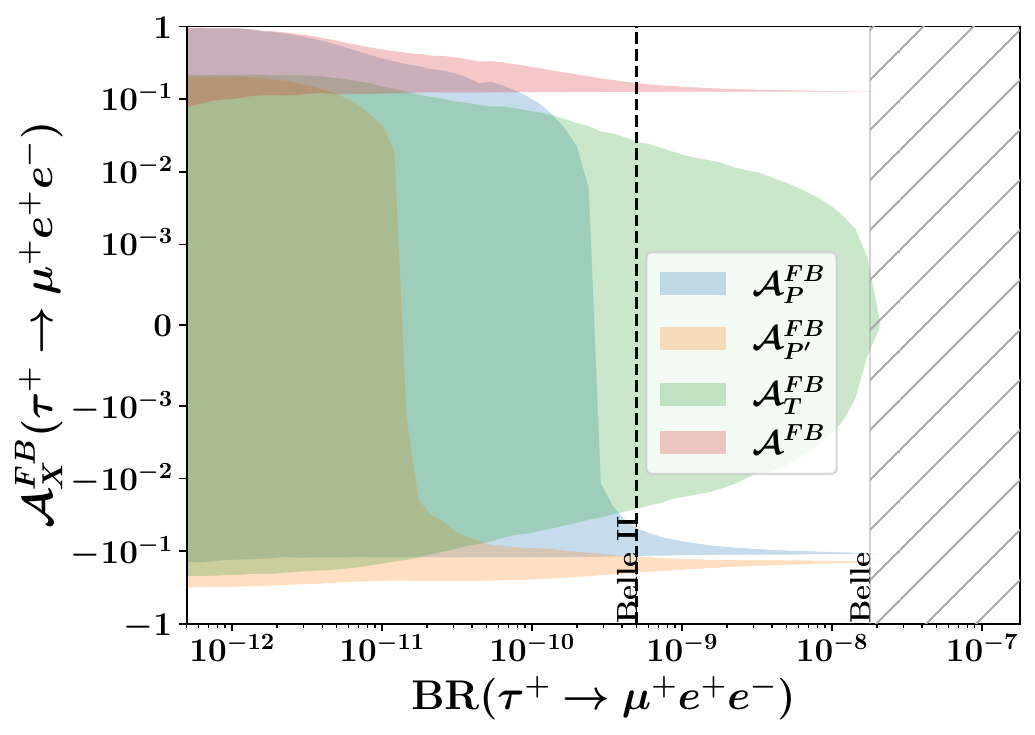}
\caption{Expected ranges of the asymmetries vs. the corresponding branching ratios for $\tau^+ \to \mu^+ e^+ e^-$ decays (on the left 
$P$, $P^\prime$ and $T$ asymmetries, while on the right the corresponding FB ones).
}
\label{fig:Asyms-BR}
\end{figure}

Applying the same analysis to the $\tau^+ \to \mu^+ e^+ e^-$ decay leads to similar conclusions, as illustrated in the panels of Fig.~\ref{fig:Asyms-BR}. In this channel, the asymmetries probe only a restricted region of the allowed parameter space. Within experimentally accessible regimes, the model's predictions occupy a small fraction of the full available asymmetry range, rendering the framework predictive (and thus falsifiable).

\section{Conclusion}

In this work~\cite{Darricau:2025rmu} we have explored the prospects for angular asymmetries in cLFV three-body decays within a minimal extension of the SM involving two sterile neutrinos. In this minimal framework, CPV phases can significantly affect correlation patterns among observables, motivating the study of a broader set of observables beyond cLFV rates alone.

We have shown that angular distributions and the associated asymmetries can help disentangle the underlying operator structure responsible for cLFV. In particular, the joint analysis between angular asymmetries and cLFV observables can significantly enhance the discriminating power between the different contributions, especially in the $\mu - e$ sector.

Even in the presence of CPV phases, which spoils simple correlation patterns between the rates of the processes, combined measurements of cLFV processes and asymmetries can still strongly constrain the parameter space. As illustrated, a simultaneous observation of BR($\mu \to 3e$)$\sim \mathcal{O}(10^{-15})$ associated with $\mathcal{A}^T \sim \mathcal{O}(20\%)$ would point towards a $\mu-e$ conversion rate in Aluminium above $10^{-14}$.

Overall, polarised asymmetries play a central role in disentangling the Wilson coefficients entering cLFV rates. Future experimental progress in $\mu$ and $\tau$ channels will therefore open a rich programme of measurements, for which the exploration of the most comprehensive set of associated observables is of paramount importance.

\bibliographystyle{ieeetr}
\bibliography{references}
\end{document}